\documentstyle{article}
\newtheorem{thm}{Theorem}
\newtheorem{prop}{Proposition}

\newtheorem{rem}{Remark}
\begin{document}
\title{Quantum groups and zeta-functions}
\author{Kimio UENO and Michitomo NISHIZAWA\\
 Department of Mathematics\\
 School of Science and Engineering\\
 Waseda University}
\maketitle
\begin{abstract}
 A $q$-analogue of the Hurwitz zeta-function is introduced through
considerations on the spectral zeta-function of quantum group
$SU_{q}(2)$, and its analytic aspects are studied via the
Euler-MacLaurin summation formula. Asymptotic formulas
of some relevant $q$-functions are discussed.
\end{abstract}

\section{Introduction}
 The left regular representation of the quantum group $SU_{q}(2)$ is,
by definition, the coordinate ring $A(SU_{q}(2))$ viewed as a
left $U_{q}(sl(2))$-module. It is known that it has the
following homogeneous decomposition.
        $$A(SU_{q}(2))=\bigoplus_{n=1}^{\infty}A_{n}$$
where each homogeneous component corresponds to the irreducible
representation of $U_{q}(sl(2))$ with spin $\frac{n-1}{2}$, and
$dim A_{n}=n^{2}$. The Casimir operator of $U_{q}(sl(2))$
        $$C=\frac{q^{-1}k^{2}+qk^{-2}-2}{(q-q^{-1})^{2}} + ef$$
acts on each homogeneous piece as a scalar operator:
\begin{equation}
        C|_{A_{n}} = \lambda_{n}
                                = \left(
                                        \frac{q^{\frac{n}{2}}-q^{-\frac{n}{2}}}
                                                {q-q^{-1}}
                                \right)^{2}.
\end{equation}\label{eqn:Cas}
Thus, according to the standard argument of differential geometry,
one can introduce a spectral zeta-function associated to the quantum
group $SU_{q}(2)$ as follows:
\begin{equation}
        Z(s:SU_{q}(2)) = \sum_{n=1}^{\infty} \frac{n^{2}}{\lambda_{n}\,^{s}},
\label{eqn:Z1} \end{equation}
which reads, substituting (\ref{eqn:Cas}) to (\ref{eqn:Z1}),
\begin{equation}
        Z(s:SU_{q}(2)) =
                (1+q^{-1})^{2s}\sum_{n=1}^{\infty}
                        \frac{n^{2}q^{ns}}{[n]_{q}\,^{2s}}
\label{eqn:Z2}\end{equation}
where $[x]_{q}=\frac{1-q^{x}}{1-q}.$\par
 These observations yields two interesting issues: The first one
is to consider the spectral geometry of quantum groups, or in general,
quantum homogeneous spaces (\cite{mas}, \cite{pod}).
 The second is to construct a $q$-analogue of zeta-functions of various
types, and to study their analytic or arithmetic aspects.\par
 We think (\ref{eqn:Z2}) tells us how to define a $q$-analogue of
zeta-functions. For instance, let us consider the multiple zeta-functions
\cite{bar2}:
        $$\zeta(s,z:\omega) = \sum_{k_{1},k_{2},\cdots k_{n} = 0}^{\infty}
                \frac{1}{(k_{1}\omega_{1}+k_{2}\omega_{2}
          +\cdots+k_{n}\omega_{n}+z)^{s}}.$$
{}From (\ref{eqn:Z2}), it seems quite natural to define a $q$-analogue
by
\begin{equation}
        \zeta(s,z:\omega,a,b:q) = \sum_{k_{1},k_{2},\cdots k_{n}=0}^{\infty}
                \frac{q^{s(a_{1}k_{1}+\cdots+a_{n}k_{n}+b)}}
                        {[k_{1}\omega_{1}+k_{2}\omega_{2}+\cdots
              +k_{n}\omega_{n}+z]_{q}\,^{s}}.
\end{equation}\par
 In this article, we restrict ourselves on the simplest case among these
$q$-multiple zeta-functions. Namely we will consider only a $q$-analogue
of the Hurwitz zeta-function (cf.\cite{kur}),
\begin{equation}
        \zeta(s,z:q) = \sum_{k=0}^{\infty}
                \frac{q^{s(k+1)}}{[k+z]_{q}\,^{s}}
\label{eqn:z1}\end{equation}
and investigate its analytic property.\par
 This paper is organized as follows: In section 2, we will show that
(\ref{eqn:z1}) has meromorphic extension to the whole $s$-plane.
The Laurant expansion at $s=0$ is particularly
important. In section 3, using the Euler-MacLaurin summation formula,
the $q$-Hurwitz zeta-function will be represented as an infinite sum
of hypergeometric functions. Section 4 treats the classical limit
($q \to 1-0$) of this zeta-function. In section 5, we will present a
functional relation satisfied by (\ref{eqn:z1}), which is regarded as
a $q$-analogue of the functional equation of the Riemann zeta-function.
In the final section, employing the results obtained in the previous
sections, we will show that an asymptotic expansion formula for the
$q$-gamma function and the $q$-shifted factorial $(q^a:q^b)_{\infty}$.
The later is related to the formula in the Ramanujan's notebook
(\cite{ber}).
A remarkable point in these formula is that Euler's dilogarithm function
appears as a principal term.\par
 The results in this paper (except the functional relation) can be
generalized to the case of the following $q$-multiple zeta-functions
\begin{eqnarray}
  \zeta_{n}(s,z:q) & = &\sum_{k_{1},k_{2},\cdots k_{n}=0}^{\infty}
     \frac{q^{s(k_{1} + k_{2} + \cdots + k_{n} + 1)}}
     {[k_{1} + \cdots + k_{n} + z]_{q}\,^{s}}\\
       & = &\sum_{k=0}^{\infty}
       \frac{{k+n-1\choose n-1} q^{s(k+1)}}{[k+z]_{q}\,^{s}}.
\end{eqnarray}
 As well as the $q$-Hurwitz zeta-function, these $q$-zeta-functions are
deeply linked to generalized hypergeometric functions and polylogarithm
functions.\par

\section{The $q$-Hurwitz zeta-function}
 Throughout this paper, we will assume that $0<q<1$ and $0<\Re z.$
The $q$-symbol $[x]_{q}$ denotes $[x]_{q}=\frac{1-q^{x}}{1-q},$ and
$\delta = \frac{2 \pi i}{\log q}.$ \par
 The Hurwitz zeta-function
 \begin{equation}
   \zeta(s,z) = \sum_{k=0}^{\infty} \frac{1}{(k+z)^{s}}
 \end{equation}
is a meromorphic function in $s.$ The pole of it is only $s=1$ and
simple. A $q$-analogue of this zeta-function is introduced by
\begin{equation}
  \zeta(s,z:q) = \sum_{k=0}^{\infty}
   \frac{q^{s(k+1)}}{[k+z]_{q}\,^{s}}.
  \label{eqn:qhur}
\end{equation}
 We call it the $q$-Hurwitz zeta-function. The right-hand side of
(\ref{eqn:qhur}) is absolutely convergent in the right half plane
$\{s|\Re s > 0 \}$, so it is holomorphic in this domain. It is easy to
show that, if $\Re s > 1$, the classical limit ($q \to 1-0$) of the
$q$-Hurwitz zeta-function coincides with the Hurwitz zeta-function.\par
 Let us consider analytic continuation of this $q$-zeta-function. By
making use of the binomial expansion theorem, we can rewrite the
right-hand side of (\ref{eqn:qhur}) to
\begin{equation}
  \zeta(s,z:q) = (q-q^{2})^{s}
    \sum_{r=0}^{\infty} \frac{(s)_{r}}{r!} \frac{q^{rz}}{1-q^{r+s}},
\end{equation}
where $(s)_{r} = s(s+1)\cdots (s+r-1).$  Since $0<q<1$, $\Re z > 0,$
the right hand side above is absolutely convergent for
\begin{equation}
  s \ne -r + \delta l,
\end{equation}
where $r$ runs over the set of non-negative integers ${\bf Z}_{\geq 0}$
and $l$ runs over the set of integers ${\bf Z}$. This implies that
$\zeta(s,z:q)$ has meromorphic continuation to the whole $s$-plane.
The poles are $s = -r + \delta l$ $(r \in {\bf Z}_{\geq 0},
l\in {\bf Z})$, and simple. It is not difficult to calculate the first
few coefficients of the Laurant expansion at $s = 0$. \par
Thus we obtain
\begin{prop}
 The $q$-Hurwitz zeta-function $\zeta(s,z:q)$ has meromorphic continuation
to the whole $s$-plane. The poles are $s = -r + \delta l$
$(r \in {\bf Z}_{\geq 0},l \in {\bf Z})$ and all simple. The Laurant expansion
at $s = 0$ reads as follows:
\begin{equation}
  \zeta(s,z:q) = \frac{\alpha_{-1}}{s} + \alpha_{0} +
       s \left\{
         \alpha_{1} -  \log \prod_{k=1}^{\infty} (1-q^{z-1+k})
         \right\} +  O(s^{2}),
\end{equation}
where
$$\alpha_{-1} = - \frac{1}{\log q}, \quad
  \alpha_{0} = \frac{1}{2} - \frac{\log(q-q^{2})}{\log q} $$
$$\alpha_{1} = -\frac{1}{12} \log q + \frac{1}{2}\log(q-q^{2})
     -\frac{1}{2}\frac{\log^{2}(q-q^{2})}{\log q} $$
and we have put $\log^{k} x = (\log x)^{k}$.
\end{prop}\par

We should notice the following fact: As mentioned before, the pole of
the Hurwitz zeta-function is only $s=1$, which is simple. In particular,
it is holomorphic at $s=0$. On the other hand, the poles of the $q$-Hurwitz
zeta-function are $s = -r + \delta l$ as pointed out in the proposition.
Namely, in the classical limit, the poles of the $q$-Hurwitz zeta-function
drastically changes. The reason for this phenomena will be explained in
section 4.

\section{The Euler-MacLaurin expansion}
 The Euler-MacLaurin summation formula is stated as follows:
 \begin{eqnarray}
   \sum_{r=M}^{N-1} f(r)  =  \int_{M}^{N}f(t)dt
       & + &  \sum_{k=1}^{n} \frac{B_{k}}{k!}
       \left\{
       f^{(k-1)}(N)-f^{(k-1)}(M)
       \right\} \nonumber\\
       & + & (-1)^{n-1} \int_{M}^{N} \frac{\overline{B}_{n}(t)}{n!}
       f^{(n)}(t)dt.
       \label{eqn:EM}
 \end{eqnarray}\par
Here $B_{k}$'s are the Bernoulli numbers and $B_{k}(t)$ are the Bernoulli
polynomial which are defined by
   $$\frac{z e^{tz}}{e^{z}-1}
      = \sum_{n=0}^{\infty} \frac{B_{n}(t)}{n!} z^{n}$$
and $B_{k}=B_{k}(0)$ ($B_{1}= -\frac12, B_{2}= \frac16, B_{4}= -\frac{1}{30},
\cdots,B_{3}=B_{5}=B_{7}= \cdots =0$). $\overline{B}_{n}(t)$ is a periodic
extension of $B_{n}(t)$ ($0 \leq t < 1$), i.e. $\overline{B}_{n}(t) =
B_{n}(t-[t]).$ \par
 Putting $f(t)= \frac{q^{(t+1)s}}{(1-q^{t+z})^{s}}$ in this summation formula,
we apply it to the study of the $q$-Hurwitz zeta-function. The k-th derivative
of $f(t)$ is expressed as
  $$f^{(k)}(t) = q^{s(t+1)} \frac{s P_{k}(q^{t+z};s)}{(1-q^{t+z})^{s+k}}
               \log^{k}q$$
where $P_{k}(x;s)$ are polynomial of $x$ defined by the following recursive
relation
  $$P_{0}(x;s) = \frac1s, \quad P_{1}(x;s)=1,$$
  $$(x-x^{2})P'_{k}(x;s) + (kx+s)P_{k}(x;s) = P_{k+1}(x;s).$$
It is found that $P_{k}(x;s)$ is a polynomial of degree $k-1$ in $x$ and
also degree $k-1$ in $s$ with positive integral coefficients, and that
$s P_{k}(1;s)= (s)_{k}, sP_{k}(0;s)=s^{k}$.\par
Putting these ingredients  into the Euler-MacLaurin summation formula,
we obtain the following expansion formula.

\begin{prop}
 When $\Re s > 0$, we have
 \begin{eqnarray}
   \zeta(s,z:q) & = &-\frac{(q-q^{2})^{s}}{s \log q} F(s,s,s+1:q^{z})
       +\frac12 \left(\frac{q-q^{2}}{1-q^{z}}\right)^{s} \nonumber \\
       & + & s \left(\frac{q-q^{2}}{1-q^{z}}\right)^{s}
       \sum_{k=1}^{m} \frac{B_{2k}}{(2k)!}
       \left(
         \frac{\log q}{q^{z}-1}
        \right)^{2k-1} P_{2k-1}(q^{z};s) \label{eqn:EMz1}\\
       & - & R_{2m}(s,z:q) \nonumber
  \end{eqnarray}
where
\begin{equation}
  R_{2m}(s.z:q) = s\int_{0}^{\infty} \frac{\overline{B}_{2m}(t)}{(2m)!}
      \left(\frac{\log q}{q^{t+z}-1}\right)^{2m}
      \left(\frac{1-q}{1-q^{t+z}}\right)^{s}
      q^{s(t+1)} P_{2m}(q^{t+z};s)dt
  \label{eqn:R}
\end{equation}
and $F(\alpha,\beta,\gamma:x)$ is a hypergeometric function.
\end{prop}\par
 We call (\ref{eqn:EMz1}) the Euler-MacLaurin expansion of $\zeta(s,z:q)$.
The appearance of a hypergeometric function in (\ref{eqn:EMz1}) is due
to the integral formula
\begin{equation}
  F(\alpha,\beta,\gamma:x)
      = \frac{\Gamma(\gamma)}{\Gamma(\beta) \Gamma(\gamma-\beta)}
        \int_{0}^{1} u^{\beta-1}(1-u)^{\gamma-\beta-1}
          (1-xu)^{-\alpha}du.
  \label{eqn:hyp}
\end{equation}\par
Now we consider the way that the $q$-Hurwitz zeta-function function is
expressed as an infinite series of hypergeometric functions.
Putting $m=1$ in (\ref{eqn:EMz1}), we get
\begin{eqnarray*}
  \zeta(s,z:q) = & -  & \frac{(q-q^{2})^{s}}{s \log q}F(s,s,s+1:q^{z})
                   + \frac12 \left(
                   \frac{q-q^{2}}{1-q^{z}}
                   \right)^{s}\\
                 & + &\frac{s}{12}
                   \left(
                   \frac{q-q^{2}}{1-q^{z}}
                   \right)^{s}
                   \left(
                   \frac{\log q}{q^{z}-1}
                   \right)
                   -R_{2}(s,z:q),
\end{eqnarray*}
where the residue term $R_{2}(s,z:q)$ is defined by (\ref{eqn:R}). Note that
\begin{equation}
  \overline{B}_{2}(t) = \frac{1}{\pi^{2}}
       \sum_{n=1}^{\infty} \frac{\cos(2\pi nt)}{n^{2}}.
\label{eqn:B2}\end{equation}
Substituting this into (\ref{eqn:R}) and making use of the integral formula
(\ref{eqn:hyp}), we obtain the following expansion formula.
\begin{thm}
  \begin{eqnarray}
   \zeta(s,z:q) & = & - \frac{(q-q^{2})^{s}}{s \log q}F(s,s,s+1:q^{z})
      + \frac12 \left(\frac{q-q^{2}}{1-q^{z}}\right)^{s}\nonumber\\
      & + &\frac{s(q-q^{2})^{s}}{2\pi i}
        \sum_{l \ne 0} \frac{1}{l(s+\delta l)}
        F(s+1,s+\delta l,s+1+\delta l:q^{z}).
    \label{eqn:EMz2}
  \end{eqnarray}
The infinite series in the right-hand side is absolutely convergent for
$s \ne -r+\delta l$ $(r \in {\bf Z}_{\geq0}, l \in {\bf Z}, l \ne 0)$ so
that it is holomorphic in this region. \label{thm:qznh}
\end{thm}

\section{Classical limit}
 Using Theorem\ref{thm:qznh}, we can achieve profound understanding for
the classical limit of $\zeta(s,z:q)$. As was mentioned before, for
$\Re s > 1$,
   $$\lim_{q \to 1-0} \zeta(s,z:q) = \zeta(s,z).$$
In this section, we will consider in detail the classical limit
of $\zeta(s,z:q)$ in the complement of this region.\par
A key to this consideration is the Gauss connection formula of a hypergeometric
function,
\begin{eqnarray}
  F(\alpha,\beta,\gamma:x)
    & = & \frac{\Gamma(\gamma) \Gamma(\alpha + \beta - \gamma)}
           {\Gamma(\alpha) \Gamma(\beta)}
      (1-x)^{\gamma - \alpha -\beta}
      F(\gamma - \alpha, \gamma - \beta, \gamma - \alpha - \beta +1:1-x)
      \nonumber\\
    & + & \frac{\Gamma(\gamma)\Gamma(\gamma - \alpha - \beta)}
            {\Gamma(\gamma - \alpha)\Gamma(\gamma - \beta)}
      F(\alpha,\beta,\alpha + \beta - \gamma +1:1-x).
  \label{eqn:con}
\end{eqnarray}
Applying this to $F(s,s,s+1:q^{z})$ and $F(s+1,s+\delta l,s+1+\delta l:q^{z})$
in (\ref{eqn:EMz2}), we get
\begin{eqnarray}
  \lefteqn{\lim_{q \to 1-0} \left\{
     \zeta(s,z:q) +    \frac{(q-q^{2})^{s}q^{-zs}}{\log q}
         \frac{\pi}{\sin \pi s}
    \right\}}    \nonumber \\
  & & = \frac{z^{1-s}}{s-1} + \frac{z^{-s}}{2}
      +\frac{sz^{-s}}{2 \pi i}
      \sum_{l \ne 0} \frac1l U(1,1-s:-2 \pi ilz)
  \label{eqn:cl1}
\end{eqnarray}
where $l$ ranges over the set of non-zero integers, and
$U(\alpha,\gamma:x)$ is a solution of the confluent hypergeometric
differential equation
  $$x \frac{d^{2}y}{dx^{2}} + (\gamma -x) \frac{dy}{dx}
       -\alpha y=0$$
satisfying, for $\Re \alpha >0$,
  \begin{equation}
    U(\alpha,\gamma:x) \sim x^{-\alpha} \quad \mbox{as} \quad x \to \infty.
     \label{eqn:U}
  \end{equation}
It is written, in terms of confluent hypergeometric functions
$F(\alpha,\gamma:x)$, as
\begin{eqnarray}
  U(\alpha,\gamma:x) & = &\frac{\Gamma(1-\gamma)}{\Gamma(1+\alpha -\gamma)}
      F(\alpha,\gamma:x) \nonumber \\
      & + & \frac{\Gamma(\gamma -1)}{\Gamma(\alpha)} e^{x} x^{1-\gamma}
      F(1- \alpha,2-\gamma:-x),
  \label{eqn:U2}
\end{eqnarray}
and has the following integral representation \cite{sla}:
\begin{equation}
  U(\alpha,\gamma:x) = \frac{1}{\Gamma(\alpha)}
     \int_{0}^{\infty} e^{-xu} (1+u)^{\gamma -\alpha -1} du.
  \label{eqn:int}
\end{equation}
Note that the right-hand side of (\ref{eqn:cl1}) is absolutely convergent
unless $s$ is a positive integer because of the asymptotic behavior
(\ref{eqn:U}).\par
Applying the Euler-MacLaurin summention formula to the Hurwitz zeta-function,
we have
\begin{equation}
  \zeta(s,z) = \frac{z^{1-s}}{s-1} + \frac{z^{-s}}{2}
             + \frac{s}{12} z^{-s-1}
             - \frac{s(s+1)}{2!}
             \int_{0}^{\infty} \frac{\overline{B}_{2}(t)}{(z+t)^{s+2}}dt.
  \label{eqn:asyz}
\end{equation}
Using (\ref{eqn:B2}) and (\ref{eqn:int}), we rewrite the integral term above.
Then we see the right-hand side of (\ref{eqn:cl1}) coincides with
(\ref{eqn:asyz}).
Namely,
\begin{equation}
  \zeta(s,z) = \frac{z^{1-s}}{s-1} + \frac{z^{-s}}{2}
             + \frac{sz^{-s}}{2\pi i}
             \sum_{l \ne 0} \frac1l U(1,1-s:-2\pi ilz).
\end{equation}\par
Thus we obtain
\begin{thm}
If $s$ is not a positive integer,
\begin{equation}
  \lim_{q \to 1-0} \left\{
     \zeta(s,z:q) + \frac{(q-q^{2})^{s}q^{-zs}}{\log q}
     \frac{\pi}{\sin \pi s}
     \right\}
  = \zeta(s,z).
\end{equation}
\end{thm}\label{thm:cll}

\section{A $q$-analogue of the functional equation for the Riemann
zeta-function}
 Let us put
 \begin{equation}
   \zeta^{*}(s,z:q) = \zeta(s,z:q) + \frac{(q-q^{2})^{s}q^{-zs}}{\log q}
                    \, \frac{\pi}{\sin \pi s}.
 \end{equation}
This is rewritten, as mentioned in the previous section, as follows:
\begin{eqnarray}
  \zeta^{*}(s,z:q) & = & \frac{1}{s-1} \left(
         \frac{q-q^{2}}{1-q^{z}}
         \right)^{s}
         \left(
         \frac{q^{z}-1}{\log q}
         \right) F(1,1,2-s:1-q^{z}) \nonumber \\
         & + & \frac12 \left(
         \frac{q-q^{2}}{1-q^{z}}
         \right)^{s}
         - S(s,z:q) + T(s.z:q)
\end{eqnarray}
where
\begin{equation}
  S(s,z:q) = (q-q^{2})^{s}q^{-zs}\frac{\Gamma(1-s)}{2 \pi i}
           \sum_{l \ne 0} \frac{1}{l}\,
           \frac{\Gamma(s+\delta l)}{\Gamma(\delta l)} e^{-2\pi ilz}
\label{eqn:S}
\end{equation}
and
\begin{equation}
  T(s,z:q) = \frac{1}{2\pi i}\left(
             \frac{q-q^{2}}{1-q^{z}}
             \right)^{s} \sum_{l \ne 0} \frac{1}{l}F(\delta l,1,1-s:1-q^{z}).
  \label{eqn:T}
\end{equation}
The right-hand side of (\ref{eqn:S}) is absolutely convergent for
$s \ne -r+\delta l$ ($r$ are positive integers and $l$ are non-zero integers)
and that of (\ref{eqn:T}) is so for $\Re s < 0$.\par
We calculate $T(s,z:q)$ under the assumption that $0 < z \leq 1$,
$\Re s < -1$, by means of the integral representation (\ref{eqn:int}) and
\begin{equation}
  \overline{B}_{1}(t) = -\frac{1}{\pi}
       \sum_{n=1}^{\infty} \frac{\sin(2\pi nt)}{n}
\end{equation}
we get
  $$T(s,z:q) = - \frac{1}{s-1} \left(
               \frac{q-q^{2}}{1-q^{z}}
               \right)^{s}
               \frac{q^{z}-1}{\log q} F(1,1,2-s:1-q^{z})
               -\frac12 \left(
               \frac{q-q^{2}}{1-q^{z}}
               \right)^{s}.$$
Thus the following theorem is deduced:
\begin{thm}
  If $0 < z \leq 1$, $\Re s < 0$ and $s \ne -r + \delta l$ ($r$ are positive
integers and $l$ are non-zero integers),
\begin{equation}
  \zeta^{*}(s,z:q) = -\frac{1}{2\pi i} (q-q^{2})^{s} q^{-zs} \Gamma(1-s)
      \sum_{l \ne 0}\frac1l\,
      \frac{\Gamma(s+\delta l)}{\Gamma(\delta l)}e^{-2\pi ilz}.
  \label{eqn:z*}
\end{equation}
\end{thm}\par
The relation (\ref{eqn:z*}) can be viewed as a $q$-analogue of the
functional equation for the Riemann zeta-function. (But it is not a
functional equation for the the $q$-Riemann zeta-function !) In fact,
{}from the Stirling formula for the gamma function,
 $$\lim_{q \to 1-0} (1-q)^{s}\frac{\Gamma(s+\delta l)}{\Gamma(\delta l)}
      = (-2\pi il)^{s}$$
is derived so that
\begin{equation}
  \lim_{q \to 1-0} \left\{R.H.S.\,of\,(\ref{eqn:z*}) \right\}
     = - \frac{\Gamma(1-s)}{2\pi i}
       \sum_{l \ne 0}\frac{(-2\pi il)^{s}}{l}e^{-2\pi ilz}.
 \label{eqn:lR}
\end{equation}
Let us introduce a generalized polylogarithm
\begin{equation}
  {\cal L}_{s}(z) = \sum_{n=1}^{\infty}\frac{e^{2\pi inz}}{n^{s}}
     \quad (\Re s > 0).
\end{equation}
Rewriting (\ref{eqn:lR}) in terms of this function, and taking
Theorem \ref{thm:cll} into account, we obtain the following functional
relation:
\begin{equation}
  \zeta(s,z) = \Gamma(1-s)\left\{
    (2\pi i)^{s-1}{\cal L}_{1-s}(z) + (-2\pi i)^{s-1}{\cal L}_{1-s}(1-z)
    \right\}.
\end{equation}
(cf.\cite{duf}, \cite{mil}). Putting $z=1$ here, we have
\begin{equation}
  \zeta(s) = 2^{s} \pi^{s-1} \Gamma(1-s) \sin(\frac{\pi s}{2})\zeta(1-s).
\end{equation}
which is nothing but the functional equation for the Riemann zeta-function
$\zeta(s)$.

\section{Stirling's formula for the $q$-gamma function and dilogarithms}
 The $q$-gamma function $\Gamma(z:q)$ is, by definition,
 \begin{equation}
   \Gamma(z:q) = (1-q)^{1-z} \frac{(q:q)_{\infty}}{(q^{z}:q)_{\infty}}.
 \label{eqn:51}\end{equation}
where
 $$(q^{z}:q)_{\infty} = \prod_{r=0}^{\infty}(1-q^{z+r}).$$\par
In this section, we will give another proof of the Stirling formula
for the $q$-gamma function, which was firstly proved by Moak \cite{moa},
and will also reprove asymptotic formulas for certain $q$-functions which
was firstly considered by Ramanujan \cite{ber}.\par
 {}From the results in section 2 and section 3, we can derive the
Euler-Maclaurin expansion for the $q$-shifted factorial$(q^{z}:q)_{\infty}$
(cf.\cite{fad}):
\begin{eqnarray}
  \log(q^{z}:q)_{\infty} & = &\frac{1}{\log q} Li_{2}(q^{z})
        - \frac{1}{12} \log q +\frac12 \log (1-q^{z})\nonumber \\
        & - &\sum_{k=1}^{m} \frac{B_{2k}}{(2k)!} \left(
        \frac{\log q}{q^{z}-1} \right)^{2k-1}P_{2k-1}(q^{z})
        + R_{2m}(z:q), \label{eqn:52}
\end{eqnarray}
where $P_{k}(x)=P_{k}(x;0)$, and
\begin{equation}
  R_{2m}(z:q) = \left. \frac{1}{s} R_{2m}(s,z:q) \right|_{s=0}
              = \int_{0}^{\infty} \frac{\overline{B}_{2m}(t)}{(2m)!}
                \left(
                \frac{\log q}{q^{t+z}-1}
                \right)^{2m} P_{2m}(q^{t+z})dz.
\label{eqn:53}\end{equation}
A remarkable point is that Euler's dilogarithm function
\begin{equation}
  Li_{2}(x) = \sum_{n=1}^{\infty} \frac{x^{n}}{n^{2}}
\end{equation}
appears in this expansion. We should note a such expansion for the
$q$-shifted factorial $(q^{z};q)_{\infty}$ had already been considered by
Ramanujan \cite{ber}.\par
We can estimate the residue term (\ref{eqn:53}) as follows: There exists
a positive constant $A$ such that
\begin{equation}
  |R_{2m}(z:q)| \leq \frac{A}{(\Re z)^{2m-1}}
  \label{eqn:55}
\end{equation}
is valid for $0 < q \leq 1$ and $0 < \Re z < +\infty$ (Moak \cite{moa}).
{}From this estimate and (\ref{eqn:51}), (\ref{eqn:52}), we can see
\begin{thm}
  As $\Re z$ tends to $+\infty$,
  \begin{eqnarray}
    \log \Gamma(z:q) & \sim & (z-\frac12) \log \left(
        \frac{1-q^{z}}{1-q} \right)
        + \log q \int_{1}^{z} \xi \frac{q^{\xi}}{1-q^{\xi}}d\xi \nonumber\\
        & + &C_{1}(q) + \frac{1}{12} \log q
        + \sum_{k=1}^{\infty} \frac{B_{2k}}{(2k)!} \left(
        \frac{\log q}{q^{z}-1} \right)^{2k-1} P_{2k-1}(q^{z}),
\end{eqnarray}
where
\begin{equation}
  C_{1}(q) = - \frac{1}{12}\log q -\frac{1}{12}\,\frac{\log q}{q-1}
           + \int _{0}^{\infty} \frac{\overline{B}_{2}(t)}{2}
           \left( \frac{\log q}{q^{t+1}-1} \right)^{2} q^{t+1}dt.
\end{equation}
This asymptotic expansion is uniformly valid for $0 < q \leq 1$,
and in particular, at $q=1$, it coincides with the Stirling formula for
$\Gamma(z)$.
\end{thm}

\begin{rem}
  The original Stirling formula holds in a section wider than $\Re z > 0$.
\end{rem}\par

As an application of the Euler-MacLaurin expansion (\ref{eqn:52}), we
give another account of the following asymptotic formula, which were
firstly proved by Ramanujan \cite{ber}, relevant to the Rogers-Ramanujan
identity:

\begin{thm}
  As $q$ tends to $1-0$,
  \begin{equation}
    \log \frac{1}{(q^2:q^5)_{\infty}(q^3:q^5)_{\infty}}
         = -\frac{\pi^{2}}{15 \log q}
           -\frac12 \log \left(\frac{5+\sqrt{5}}{2}\right) + O(\log q),
  \label{eqn:58} \end{equation}
  \begin{equation}
    \log \frac{1}{(q:q^5)_{\infty}(q^4:q^5)_{\infty}}
         = -\frac{\pi^2}{15\log q}
           -\frac12 \log \left(\frac{5-\sqrt{5}}{2}\right) + O(\log q).
  \label{eqn:59}\end{equation}
\end{thm}\par

{}From (\ref{eqn:52}), we have
\begin{eqnarray}
  \log(q^a:q^b)_{\infty} & = & \frac{1}{b \log q} Li_{2}(q^a)
       + \frac12 \log (1-q^a) - \frac{b}{12}\frac{q^{a} \log q}{1-q^a}
       \nonumber \\
       & - & R_{2}(a,b:q),
  \label{eqn:510}
\end{eqnarray}
where
\begin{equation}
  R_{2}(a,b:q)
     = \int_{0}^{\infty} \frac{\overline{B}_{2}(t)}{2}\,
       \frac{q^{at+b}(a \log q)^{2}}{(1-q^{at+b})^{2}}dt.
\end{equation}
Since $\overline{B}_{2}(t)$ is a periodic extension of $B_{2}(t) = t^{2}
-t + \frac16 \mbox{ } (0 \leq t <1)$, we get
\begin{equation}
  \lim_{q \to 1-0} R_{2}(a,b:q) = \frac{1}{12\alpha}
        + \frac12 \log \left\{
        \prod_{n=0}^{\infty} \left(
        \frac{n+\alpha}{n+\alpha+1}
        \right)^{2n+2\alpha +1} e^{2}
        \right\},
\end{equation}
where $\alpha = \frac{a}{b}$. Here using the Weierstass product formulas
for the gamma function and Barnes' $G$-function \cite{bar1}
  $$\prod_{n=1}^{\infty} \left\{
    \left(
    1+\frac{\alpha}{n}
    \right) e^{-\frac{\alpha}{n}}
    \right\}
   = \frac{e^{-\gamma \alpha}}{\Gamma(\alpha+1)},$$
 $$\prod_{n=1}^{\infty} \left\{
  \left(
  1+\frac{\alpha}{n}
  \right) e^{-\alpha + \frac{\alpha}{2n}}
  \right\}
  = (2\pi)^{-\frac{\alpha}{2}}
     e^{\frac{\alpha(\alpha +1)}{2}+\gamma \frac{\alpha}{2}}
     G(\alpha +1),$$
where $\gamma$ is the Euler constant, we have
  \begin{equation}
     \lim_{q \to 1-0} R_{2}(a,b:q) = \frac{b}{12a} - \frac12 \log(2\pi)
       - \frac{a}{2b} - \left(\frac{a}{b}-\frac12 \right) \log\frac{a}{b}
       + \log \Gamma(\frac{a}{b}).
      \label{eqn:513}
   \end{equation}
Also we have the following asymptotic formula (Lewin \cite{lew})
  \begin{equation}
    \frac{1}{b\log q}Li_{2}(q^{a})
       = \frac{\pi^{2}}{6b\log q} - \frac{a}{b} \log a
         - \frac{a}{b} \log(-\log q) + \frac{a}{b} + O(\log q).
    \label{eqn:514}
  \end{equation}
Substituting (\ref{eqn:513}), (\ref{eqn:514}) in the classical limit
of the right-hand side of (\ref{eqn:510}), we obtain the following
theorem.

\begin{thm}
 As $q$ tends to $1-0$,
 \begin{eqnarray}
  \log (q^a:q^b)_{\infty} = \frac{\pi^{2}}{6b\log q} & + &
     \left(
       \frac12 - \frac{a}{b}
     \right)
     \left\{
      \log (-\log q) + \log b
     \right\}
     + \frac12 \log(2\pi) \nonumber \\
     & - & \log \Gamma(\frac{a}{b}) + O(\log q).
   \label{eqn:515}
 \end{eqnarray}
\end{thm}\par

As a corollary of the theorem, we have
\begin{equation}
  \log \left\{
    (q^a:q^{a+b})_{\infty}(q^b:q^{a+b})_{\infty}
    \right\}
  = \frac{\pi^{2}}{3(a+b)\log q}
       + \log \left\{2 \sin
       \left( \frac{a}{a+b}  \pi \right)
       \right\} + O(\log q).
  \label{eqn:516}
\end{equation}
where we have used the reciprocal identity of the gamma function.
 Putting $a=2$, $b=3$ (resp. $a=1$, $b=4$), we get (\ref{eqn:58})
(resp. (\ref{eqn:59})). \par
\newpage
\noindent{\large\bf Acknowledgement.}\par
The first author thanks Professor J.Lukierski and Professor Z.Popowicz
for their hospitality during the Karpacz Winter School. \par
The first author is partially supported by Grant-in-Aid for Scientific
Research on Priority Area 231 ``Infinite Analysis'', and by Waseda
university Grant for Special Research Project 94A-221.\par

\end{document}